\documentclass[10pt,aps,prx,twocolumn,superscriptaddress,
 amsmath,amssymb,amsfonts,
 floatfix
]{revtex4-2}
\usepackage{graphicx}
\usepackage{bm}
\usepackage{appendix}
\usepackage[dvipsnames]{xcolor}
\usepackage{hyperref}
\hypersetup{
    colorlinks=true,
    citecolor=Green,
    linkcolor=Red,
    urlcolor=NavyBlue
}
\usepackage[capitalise]{cleveref}
\usepackage[italicdiff]{physics}
\usepackage{relsize}
\DeclareUnicodeCharacter{2212}{-}

\usepackage{ulem}

\begin{abstract}
"Stripes" - meaning unidirectional charge-density-waves, sometimes (but not always) accompanied by spin-density-waves with twice the period - are now known to arise in broad swathes of the cuprate phase diagram, and appear as a strong ordering tendency in numerical studies of Hubbard-like models of highly correlated electron systems. Jan Zaanen's work played a seminal role in predicting their existence, and exploring their possible significance. They are {\it not} related to any weak-coupling physics associated with some form of Fermi-surface nesting. And whether one likes them or not, they are surprisingly difficult to avoid;  in the Hubbard model, for example, they often appear as an alternative order that can out-compete the otherwise favored $d$-wave superconductivity. 
\end{abstract}

\begin{document}

\title{The significance of ``stripes'' in the physics of the cuprates, the Hubbard model, and other highly correlated electronic systems}

\author{Thomas P. Devereaux}
\email{tpd@stanford.edu}
\affiliation{Stanford Institute for Materials and Energy Sciences,
SLAC National Accelerator Laboratory, 2575 Sand Hill Road, Menlo Park, CA 94025, USA}
\affiliation{
Department of Materials Science and Engineering, Stanford University, Stanford, CA 94305, USA}
\affiliation{
Geballe Laboratory for Advanced Materials, Stanford University, Stanford, CA 94305, USA}

\author{Steven A. Kivelson}
\email{kivelson@stanford.edu}
\affiliation{
Department of Physics, Stanford University, Stanford, CA 94305, USA}

\maketitle
{\bf Stripes:}  One of Jan Zaanen's most influential works was his prediction, in collaboration with Gunnarson (ZG) \cite{zg}, of the existence of unidirectional spin- and charge-density-wave order (spin and charge ``stripes'') in the cuprate high temperature superconductors. The calculations, and others of the time \cite{Schulz1,Schulz2,Machida,Poilblanc,Littlewood}, involved were simple - essentially Hartree-Fock.  Their significance for the physics of the cuprates (and, more generally, for the physics of highly correlated materials) was at first unclear - and indeed, for many years following this work, those who considered charge-density-wave (CDW) phenomena to be in any way broadly relevant to cuprate physics constituted a fringe minority in the field.  (Jan was often a whole-hearted member of one or another heterodox fringe!) 

As a matter of principle, superconductivity is straightforward to understand on the basis of the well established physics of simple metals;  the Cooper instability occurs for arbitrarily weak coupling, and the resulting superconducting state can be understood in this limit on the basis of one or another extension of BCS mean-field theory.  Of course, ``high temperature superconductivity'' cannot possibly be a weak-coupling phenomenon, and indeed the ``normal'' state out of which the superconducting state emerges is anything but a normal metal.  Thus, there remain significant open issues concerning the optimal way to view the ``mechanism'' of superconductivity in the cuprates, and especially concerning all the most important ingredients that make quantitative contributions to determining the transition temperature, $T_c$.  None-the-less, there is a broad (if not always well articulated) consensus that the $d$-wave superconductivity can be understood - by adiabatic continuity if needed - on the basis of the BCS gap equation with a strongly $\vec k$-dependent but at most weakly retarded repulsion between electrons that reflects the presence of {\it short-range} antiferromagnetic correlations \cite{scalapinocommonthread,Keimer2015}.

By contrast, in more than 1D, CDW order does not generically arise at weak coupling.  It requires intermediate scale interactions, $V \gtrsim E_F$.  The fact that CDW order can arise in 1D, through a weak-coupling Peierls instability, has led to the widely accepted fallacy that Fermi surface nesting is the essential driver of CDW order even in higher dimensions.  However, recent studies have documented clearly that CDW ordering is not generically responsive to details of the Fermi surface structure, and depends in at least as large measure on the strength, and $\vec k$-dependence of the electron-phonon coupling and similar considerations \cite{mazin,eiter,esterlis}.  In the cuprates, in particular, the stripe order that is seen arises from doping an antiferromagnetic (AF) insulator. Indeed, in the scenario of ZG (which is not quite what happens in the cuprates), light doping converts the AF insulator into a striped insulator, with no intervening metallic phase at all.

For a period in the history of the cuprates, the issue of what happens upon light doping of a ``Mott insulating'' AF was thought to be the key theoretical problem that needed addressing.  Attempts to address this problem produced a cornucopia of imaginative proposals that - if not in the end of obvious relevance to the cuprates - have proven to have enormous implications for the field in general \cite{PWA1987,Emery1987}.  One proposal, by Emery and one of us \cite{ek}, was that, absent long-range Coulomb interactions, a doped AF insulator is intrinsically unstable to phase separation into hole-rich metallic regions and undoped AF insulating regions.  Inclusion of weak long-range Coulomb interactions was shown \cite{lowe} to generically lead to metallic stripes.  Somewhat later, in studies of the Hubbard model at intermediate couplings, White and Scalapino \cite{ws1,ws} discovered a tendency of dilute holes to cluster into stripes - much like the stripes of ZG but with a higher density so that the stripes are ``partially filled'' and hence conducting.  As was stressed by Zaanen
\cite{ZAANEN1998}, although these three proposals are distinct in some ways, they all basically involve a form of local phase separation into mesoscale doped and undoped regimes.  They are, in this sense, driven largely by spatially local physics and are distinctly unrelated to particular nested portions of any putative Fermi surface.

In the remainder of this note we discuss - in qualitative terms - some of the progress that has been made in understanding the role of stripes in the physics of the cuprates, the Hubbard model, and other interesting strongly correlated systems.  We lump these together not because we believe that the Hubbard model is a reasonable microscopic model of the cuprates. (It is not.)  We view the Hubbard model as a particularly simple highly correlated material that is ideal for theoretical investigation, just as BSCCO is the cuprate most suitable for surface spectroscopy (e.g. angle-resolved photoemission (ARPES) and scanning tunneling microscopy (STM)) and LSCO and YBCO are ideal for diffraction studies (i.e. neutrons and X-rays).  The underlying principle is that while there are notable microscopic differences between all these systems, they have enough in common that studying one teaches us about the others as well.  We will in particular emphasize the far from a priori obvious observation that - in the range of parameters in which high temperature superconductivity arises - stripe ordering tendencies of some sort appear to be more or less ubiquitous.

{\bf Stripes in  Hubbard-Like Models:} Although a large body of serious numerical study has been devoted to the two-dimensional Hubbard model and its variants - three-band, two-band, single-band, and $t-J$ models - the challenge of understanding a generic phase diagram has yet to be met \cite{annurev}. Although overall agreement at half-filling (antiferromagnetic insulator) and at large hole and electron dopings (weakly correlated metal) has been obtained, details about how each phase evolves into the others remain controversial, largely due to the plethora of phases that have all been found to have approximately similar ground state energies. 
Therefore, small differences in theoretical approaches - such as the use of different numerical techniques, cluster shapes and sizes, boundary conditions, parameter choices, number of variational parameters, and so on, seem to matter greatly \cite{rts,annurev,gullrev,Corboz1,Simons,Corboz,review,4leg-7,4leg-5,6leg-4,AG,Shengtao}. Symptomatic of this problem is that the energies inferred for a variety of distinct candidate ground states extracted using various numerical  methods, such as density matrix renormalization group (DMRG), 
infinite projected entangled-pair state (iPEPS), variational Monte Carlo (VMC), and exact diagonalization (ED), for example, differ in energy density per site by of order $10^{-2}t$ or less, where $t$ the nearest-neighbor hopping \cite{Corboz1,Simons,Corboz}.

DMRG is a particularly attractive approach to this problem.  It is a variational algorithm based on matrix product states in which
the wave function is efficiently represented by a trace over
the product of tensors.  It has probably provided the most detailed window into the important ground-state ordering tendencies at short to intermediate length scales.  
Based on the control available for calculating interacting systems in one-dimension, DMRG has been used to study 2D systems (typically on cylinders) by mapping the system onto a 1D problem with long-ranged interactions \cite{review}. Crucially, the accuracy of the DMRG results on any finite cluster can be systematically improved by increasing the bond dimension $D$ of the tensors, or equivalently, the number of states $m$ kept during the truncated part of the calculation.

To obtain sufficient confidence in the numerical results of the two-dimensional DMRG, one requires that the results do not change appreciably as $D$ increases beyond a certain size, as the boundaries change, or as the size of the cluster increases. The main obstacle to overcome is that the number of states $m$ kept must increase exponentially with the width of the system. 
At present, this more or less limits the size of systems that can be adequately addressed to ladders under width-6 for Hubbard models, and width-8 for $t-J$ models. 

Moreover, 
for any finite $D$ and finite number of DMRG ``sweeps,'' it is not guaranteed that the DMRG algorithm will achieve the true ground state: the number of states may be too small to represent the wavefunction accurately, and if the entanglement grows it can get stuck in metastable states and never reach the ground state. 
For example, many DMRG calculations yield results that break symmetries that are expected on general theoretical grounds to remain unbroken, most particularly the $SU(2)$ spin rotational symmetry. 
Nevertheless, due to the growth in memory-on-a-chip, GPU acceleration, and further algorithm developments, the amount of states $m$ kept during the DMRG simulations has increased greatly, 
giving reason to hope that the plethora of contradictory results that characterize the assertions made in much of the present literature will resolve into consensus in the near future \cite{Corboz1,Simons,Corboz,review,6leg-4,AG,Shengtao}. 

One of the phases that appears to be ubiquitous as long as it can be fit into the cluster geometry are stripe phases, very much like those ZG forecasted. Although the initial finding from DMRG of stripes in the $t-J$ model for width-4 and shortly thereafter width-8 cylinders  \cite{ws1,ws} was met with skepticism, the old quote "everywhere is within walking distance if you have the time" was realized: stripes now regularly appear in most calculations. Care to minimize Friedel charge density oscillations from boundaries, sensitivity to parameter choices (such as finite $t'$), as well as analysis of behavior extrapolated to large cylinder lengths and larger number states kept represent current DMRG challenges that continue to be addressed via increased computational firepower. 

To date, the bulk of DMRG studies involve values of $U$ on the order of the bandwidth (i.e. $8t \leq U \leq 12t$) or corresponding values of $J$ (i.e. $t/3 \leq J \leq t/2$) such that at half-filling the ground state is N\'eel ordered and insulating. For finite hole doping, stripes appear in both $t-t'-J$ and Hubbard models. Refs. \cite{annurev} and \cite{gullrev} review the state-of-the-art at the time of publication in greater detail, which can be summarized as follows:
\begin{itemize}
    \item Small width-ladders show dominant CDW correlations and Luther-Emery phases. 
    \item As the width of the ladder is increased, a more diverse array of behaviors is seen.
    \item A strong sensitivity to next nearest neighbor $t'$ can dramatically alter the wavelength and magnitude of CDW modulations.
    \item A strong particle-hole asymmetry is seen, with a tendency towards stronger CDW correlations for hole-doping as opposed to electron doping. 
    Specifically, in Hubbard models with $t' <0$ (to mimic the cuprate band-structure), increasing $|t'|$ weakens CDW correlations more effectively for electron than hole doping. 
    \item
    Not surprisingly, superconducting correlations appear stronger whenever CDW correlations are weaker.
\end{itemize}
For example, in width-2 cylinders, where detailed control can be obtained, a CDW wavevector $Q=2k_F=2\pi x$, with $x$ the concentration of doped holes, has been convincingly shown to possess a Luther-Emery liquid groundstate \cite{tJ-lu-2}. 

For width-4 cylinders, a sensitive dependence on $t'$ can change the wavevector of CDW order by a factor of 2, transitioning between filled or half-filled stripes \cite{4leg-1,4leg-2,4leg-3,4leg-4,4leg-5,4leg-6,4leg-7,4leg-8}. Concomitantly, the magnitude of charge density variation generally decreases for increasing $|t'|$, 
with long-distance correlations still 
consistent with characterizing the phase as a Luther-Emery liquid. 

For width-6 cylinders, the evolution of the quantum phases is surprisingly sensitive to both $t'$ and $x$ \cite{Hager,gong,6leg-4,6leg-5}. 
Without $t'$ the ground state forms an insulating stripe phase with 
unidirectional “2/3-filled” charge stripes and mutually commensurate spin stripes, but short-range (exponentially localized) superconducting correlations. In stark contrast to width-4 cylinders, for finite $t'<0$ the ground states remain fully insulating while a robust $d$-wave phase, similar to the one observed in the Hubbard model on four-leg cylinders is found for $t'>0$.
(A particle-hole transformation relates the Hubbard model with a given value of $x$ and $t'$ to the model with $x \to -x$ (i.e. electron-doped) and $t'\to -t'$;  thus, the hole-doped model with $t'>0$ is typically considered a model of electron-doped cuprates.)

For width-8 cylinders, to date results have been obtained in the more forgiving $t-t'-J$ model, but with conflicting results \cite{ST-1,ST-2,donna,chen2023}. 
Refs. \cite{donna,chen2023} observed evidence for a $d$-wave SC on the hole-doped side at the optimal 1/8 doping, while Refs. \cite{ST-1,ST-2} drew qualitatively similar conclusions as from width-6 cylinders (i.e. CDW rather than SC) for both positive and negative $t'$. 
Clearly, while DMRG progress has been proceeding at a quick pace, results are not yet conclusive as to whether (and in what range of parameters) the Hubbard or $t-J$ model is dominantly superconducting or CDW ordered  in the 2D limit. 

One can also approach the problem from finite temperature using  
determinant quantum Monte Carlo (DQMC). Here the infamous fermion minus sign problem precludes reaching as low temperatures as would be desired. Already, early results showed a tendency to form $d_{x^2-y^2}$ pairs, but no evidence for a finite temperature phase transition could be identified \cite{White}. More recently, much larger scale DQMC calculations have reached lower temperatures (of order half the exchange energy $J/k_B$) and given a more complete handle on the way the various ordering tendencies grow upon cooling.  Fluctuating stripe orders, similar texturally to those observed by DMRG, have been found 
to be easily identifiable at temperatures as high as $J/k_B$ 
\cite{edwin1,edwin2}.  These calculations also show that the dominant SC susceptibility is for $d$-wave pairing and is peaked at $q=0$ (i.e. there is no tendency to form a pair-density-wave), its growth with decreasing $T$ in the accessible range is only modest \cite{JSPJ}.  

Thus, DMRG and DQMC calculations are moderately consistent concerning the prevalent short-range correlations.  This is not surprising since short-range correlations are what determine the energy, and so are generally more robust than subtle long-range correlations. It is clear that there is a strong local tendency to $d$-wave superconducting pairing, as well as strong tendencies to CDW order - mostly unidirectional but with a preferred periodicity that seems to vary in complicated ways with changes of parameters.  This supports a picture of the stripes as locally phase separated regions where the itineracy of the doped holes is not obstructed by local AF order, while the exact pattern in which these regions organize is determined by smaller energies in the problem.  There also are clear signatures of short-range AF correlations peaked at $(\pi,\pi)$, although in many cases there is a large spin-gap and, correspondingly, the AF correlations are very short-range. Presumably, at long distances, and in the 2D limit where true broken symmetry states arise, one or the other of these orders likely dominates and suppresses the others.  It certainly seems reasonably clear from the bulk of DMRG and DQMC calculations that the CDW ordering wins over superconductivity in the case of $t'\le 0$. 

Results obtained with other numerical approaches lead to similar conclusions concerning the relevant ordering tendencies, and also reflect the same ambiguities about which orders win under which circumstances as does DMRG.  For instance, 
dynamical mean field cluster methods \cite{Maier,Gull,gullrev}, where the momentum structure of the Hubbard model self-energy is approximated by $N_c$ basis functions which retain the full frequency dependence, as well as constrained path quantum Monte Carlo along with DMRG in the presence of applied superconducting potentials at the leads \cite{Xu2024} have provided evidence for $d-$wave pairing.  On the other hand, when the cluster size and shape are appropriately chosen, stripe order of various types also shows up strongly in such cluster mean field approaches \cite{Maier2022}. Along with the conflicting results mentioned above from the width-8 DMRG results, this makes it clear that the final word has not yet been uttered.

This leads to the question, given that the cuprates manage to be robust superconductors, ``what is missing in the Hubbard model"? A
tongue-in-cheek answer might be ``consensus.'' 
More seriously, there is little doubt that (in common with the cuprates) antiferromagnetism, $d$-wave SC, and the stripes that ZG envisioned are
major players in
all Hubbard-like models at intermediate couplings.  However, 
the intertwined nature of these ordering tendencies (and what determines the delicate balance between them) is still yet to be unraveled. 

{\bf Stripes in perspective:}
What relevance does all this discussion of stripes have for the ``essential'' physics of the cuprates?  This is  the sort of question Zaanen loved to debate.  We do not pretend to have a definitive answer to this question - or even a precise definition of what it means.  But in honor of his enthusiasm for such discourse, we thought to end with a few thoughts on this topic.

For many years after experimental evidence of stripe order was first discovered \cite{Tranquada1995}, it was widely believed that CDW order would prove to be a peculiarity - either to do with the special LTT crystal structure of LNSCO where it was first discovered, or of the commensurate $x=1/8$ doping. It is clear, by now, that unidirectional CDW correlations (i.e. with correlation lengths long compared to the period) are a moderately ubiquitous feature of at least the hole-doped cuprates. Moreover, as more discriminating experiments are carried out, the range of doping, $x$, over which identifiable CDW correlations are reported has steadily increased.  Whatever the relation between CDW order and high temperature superconductivity, it is not something that can simply be avoided.

However, that leaves several points of perspective unaddressed, among them:  1) Do all the density-wave-orders seen experimentally reflect a single ordering tendency, or are various forms of spin and charge density wave - maybe with rather different origins - being conflated?  2)  Are the observed density-waves, in particular, related to the corresponding orders that have been identified in one or another theoretical treatment of the purely electronic models we have been discussing?  (Note, that in other contexts, when one discusses CDW order, one usually invokes the effects of strong electron-phonon couplings - couplings that certainly play some role in the corresponding physics of the cuprates as well.)  3) Are these ``important'' phenomena, or simply just another material-specific ``microscopic detail'' that we would do better to ignore?

{\underline {Concerning 1}}:  The majority of the CDW order reported in the cuprates \cite{Tranquada1995,Giacomo,Comin,XRD,Riccardo,Fabio,Hayden,STM1,STM2,STM3,STM4,STM5} has several features in common - which are shared with much of the CDW order seen in the model calculations we have discussed. At least in the superconducting range of $x$, the CDW order is always unidirectional with ordering vector more or less aligned with the Cu-O bond direction, it is strongest in a range of underdoping where $x \sim 1/8$, and it does not fully gap the electronic spectrum (i.e. these are ''partially filled'' conducting stripes).  However, while the period of the CDW order is comparable in different cuprates, neither its magnitude nor its doping ($x$) dependences are quite the same.  Moreover, in some cases CDW order promotes coexisting SDW order with twice the period of the CDW order (as in the case of the Hartree-Fock stripes of ZG), while in other cases the CDW and SDW orders are mutually incommensurate, and seem to compete sufficiently strongly that they fail to coexist.  While it is possible - simply from a Landau-Ginzburg perspective - to provide \cite{differentstripes} a unified description of these various behaviors, it is also plausible that they reflect more profound differences in the physics in different cuprate families.  This is an important topic for further investigation. 

{\underline {Concerning 2}}: 
Going into more details, the evidence from experiments on charge order \cite{Tranquada1995,Giacomo,Comin,XRD,Riccardo,Fabio,Hayden,STM1,STM2,STM3,STM4,STM5} also gives a surprising material dependence that does not seem to be reflected as small parameter changes in Hubbard model calculations. While the off-axis charge transfer layers admit steric forces that relate to oxygen ordering on surfaces and in the bulk, the presence or absence of oxygen ordering can be separated from CDW-like order via temperature and doping dependences. To date, the observance of a decreasing wavevector of CDW-like order with increased hole doping in BSCCO and YBCO-families has no clear analog from DMRG on the Hubbard model. 

It is not clear exactly what can be drawn from this observation, but presumably the phonon degrees of freedom could be playing a critical role. After all, they play dominant roles in various stripe charge orders observed in nickelates \cite{nickelates}, manganites \cite{manganites1}, and other transition metal oxides \cite{manganites}, and evidence is ample that in-plane bond-stretching modes that transfer charge between transition metal and ligand oxygens are exactly the sort of displacements that may couple strongly to the quasi one-dimensional CDW modulations \cite{BS1,BS2,BS3,BS4,BS5,BS6}. 

Early on, attributing ARPES dispersion anomalies ("kinks") to coupling to some sort of bosonic mode in the cuprates, be it spin fluctuations or phonons, was a realm of lively discussion. Zaanen eagerly took what was at the time a somewhat unorthodox view that the kinks were due to phonons. His willingly "retro" view led to a body of collaborative work with one on us.

ARPES \cite{ARPES} has delivered strong empirical support for the claim that the electron-phonon (EP) coupling may be quite consequential \cite{Cukr}. At very low doping the screening is poor and one is dealing with the physics of isolated electric charges (the carriers) moving around in an ionic crystal giving rise to strong polar EP interactions \cite{non1,non2}.  These are actually identifiable in experiment, in the form of polaron features imprinting on the ARPES at low dopings \cite{Shen}. Strikingly, when the quasiparticles restore in the superconductor the fingerprints of strong EP physics can be identified \cite{ARPES,Cukr,aspects,Hash,Hash1}. However, when the system switches to the overdoped Fermi-liquid-like regime, these suddenly disappear, consistent with the expectation of a rather moderate EP coupling in good metals \cite{rapid}. 

Unfortunately, calculations possessing the same accuracy on Hubbard models that include electron-lattice interactions on the same footing as electron-electron interactions are limited to perhaps oversimplified model Hamiltonians. This remains an interesting area for future study.

{\underline {Concerning 3}}:  
There are at least two serious reasons to argue that the CDW order in the hole-doped cuprates is ``weak,'' and therefore possibly not central to the physics on energy and temperature scales of order $T_c$:  1) Firstly, the scattering peaks seen in hard X-ray diffraction are very weak, meaning that the magnitude of the lattice displacements induced by the CDW order are exceedingly small \cite{Riccardo,Comin,XRD}.  To quantify this, we can compare the cuprates with  the rare earth tritellurides (RTe$_3$), which have CDW transition temperatures comparable to the onset temperatures of stripe signals in the cuprates \cite{IRF,RET}.  In RTe$_3$, the CDW peaks have an intensity of order 10$^{-3}$ times the crystalline Bragg peaks, while in the cuprates the CDW peaks are more like 10$^{-6}$ times smaller.  (This suggests lattice displacements that are $30 \sim \sqrt{1000}$ times smaller in the cuprates.)  Secondly, (again in contrast with the situation in RTe$_3$) no clear evidence of reconstruction of the electronic dispersion (observable in ARPES) associated with the onset of stripe order has been reported in the cuprates, suggesting that in some sense its effect on the important low energy degrees of freedom is unexpectedly subtle. 

Neither of these objections is as unambiguous as they may at first seem.  The energy scales involved in the CDW ordering are manifestly comparable to those for superconductivity.  This is reflected both in the fact that the CDW onset temperatures are comparable to (or possibly somewhat higher than) the typical scales of $T_c$, and in the fact that - especially in the neighborhood of $x=1/8$, CDW order manifestly suppresses long-range superconducting phase coherence.  The interrelation between superconductivity and CDW order is 
also manifest in experiments in which superconductivity is suppressed by application of high magnetic fields \cite{Gerber2015,Chang2016,Lee2016} and/or strain \cite{LeTacon,LeTacon1,Choi2022,Guguchia}; an increase of one order always occurs at the expense of the other.

Thus, it is plausible that
{the CDW order is indeed central to the physics of the cuprates across the phase diagram, 
even while} the small lattice displacements imply a secondary role for electron-phonon coupling in the mechanism of CDW formation.  This, in turn, would imply that the magnitude of the ionic displacements does not provide a reasonable estimate of the condensation energy.  The ARPES data comes primarily from experiments on single and double-layer BSCCO, where (possibly due to strong disorder effects) the CDW correlation lengths inferred from X-ray diffraction \cite{Riccardo,Comin,XRD} or STM \cite{STM1,STM2,STM3,STM4,STM5}  are no longer than the CDW period;  possibly this is responsible for obscuring signatures that would otherwise be observable. 

{\bf Coda:}
For Jan, the ``high temperature'' superconductivity of the cuprates - indeed even the superconductivity itself - were not of primary interest.  Instead, he insisted that what is  essential  is the abnormal  ``normal state'' and the   unprecedented complexity and richness of the phase diagram that emerges from it at low $T$.
He  often waxed poetic about  the inner workings of the cuprates - what he called ``nature's quantum computer.''
In his view, the phenonena and epiphenomena seen in the cuprates  constitute a wake-up-call that much of what has long been the accepted theoretical structure for understanding  electronic fluids in quantum materials  is fundamentally flawed - many of its successes more representative of confirmation bias than of objective truth. Rather than summarizing our own perspective, we close with some words from Jan on the cuprates and his vision of their impact on modern physics. 

{\it A pursuit that started as a rather folkloristic affair rooted in 1950's sentiments has turned thanks to the power of condensed matter experimentation in 2019 into a pursuit studying the deepest fundaments of physics itself, with the promise to demonstrate that there are completely different realms of hitherto unidentified  forms of macroscopic quantum physics that may have even ramifications for the understanding of quantum gravity.}

\begin{acknowledgments}
We acknowledge discussions with Hong-Chen Jiang, Shengtao Jiang, Andy Millis, and Richard Scalettar.
This work was supported by the U.S. Department of Energy (DOE), Office of Basic Energy Sciences,
Computational work was performed on the Sherlock cluster at Stanford University and on resources of the National Energy Research Scientific Computing Center, supported by the U.S. DOE, Office of Science, under Contract no. DE-AC02-05CH11231.
\end{acknowledgments}

\bibliography{main}

\end{document}